# Pressure-induced recovery of the Fermi-liquid state in the non-Fermi liquid material $U_2Pt_2In$

P. Estrela[a], A. de Visser[a,*], T. Naka[a,b], F.R. de Boer[a] and L.C.J. Pereira[c]

[a] *Van der Waals-Zeeman Institute, University of Amsterdam,*
*Valckenierstraat 65, 1018 XE Amsterdam, The Netherlands*
[b] *National Research Institute for Metals, 1-2-1 Sengen, Tsukuba, Ibaraki 305-0047, Japan*
[c] *Department of Chemistry, Technological and Nuclear Institute,*
*Apartado 21, 2686-953 Sacavém, Portugal*

In the study of non-Fermi-liquid (NFL) phenomena in correlated metals, $U_2Pt_2In$ is of special interest as it is one of the rare stoichiometric (undoped) materials that show NFL behaviour at ambient pressure. Here we report on the stability of the NFL phase with respect to hydrostatic pressure ($p \leq 1.8$ GPa). Electrical resistivity data under pressure, taken on a single-crystalline sample for a current in the tetragonal plane, show that $T_{FL}$, i.e. the temperature below which the Fermi-liquid $T^2$-term is observed, increases with pressure as $T_{FL} \sim (p-p_c)$, where $p_c \approx 0$ is a critical pressure. This provides strong evidence for the location of $U_2Pt_2In$ at an antiferromagnetic quantum critical point.

Keywords: Non-Fermi liquid, quantum critical point, high-pressure, resistivity, $U_2Pt_2In$

*Author to whom correspondence should be addressed:
   Dr. A. de Visser
   Van der Waals-Zeeman Institute, University of Amsterdam
   Valckenierstraat 65, 1018 XE Amsterdam
   The Netherlands
   Phone: +31-20-5255732; Fax: +31-20-5255788
   E-mail: devisser@science.uva.nl

Among the $U_2T_2X$ compounds, where T is a transition metal and X is In or Sn, $U_2Pt_2In$ takes a special place, because it is a non-ordering heavy-electron compound with a strongly renormalised quasiparticle mass ($c/T = 0.41$ J/molU-$K^2$ at $T = 1$ K) [1]. Moreover, it shows pronounced non-Fermi liquid (NFL) behaviour. The NFL properties are summarised by: (i) the specific heat varies as $c(T) \sim -T\ln(T/T_0)$ over almost two decades of temperature ($T = 0.1$-$6$ K) [2], (ii) the magnetic susceptibility [3] shows a weak maximum at $T_m = 8$ K for a magnetic field along the $c$ axis (tetragonal structure), while it increases as $T^{0.7}$ when $T \rightarrow 0$ for a field along the $a$ axis, and (iii) the electrical resistivity obeys a power law $T^{\alpha}$ with $\alpha = 1.25 \pm 0.05$ ($T < 1$ K) and $0.9 \pm 0.1$ ($T \rightarrow 0$), for the current along the $a$ and $c$ axis, respectively [3,4]. Notice that muon spin relaxation experiments [4] have demonstrated the absence of (weak) static magnetic order at least down to 0.05 K. In order to investigate the stability of the NFL phase with respect to pressure, we have carried out a high-pressure transport study on single-crystalline $U_2Pt_2In$ [4]. The electrical resistivity, $\rho(T)$, was measured for a current, $I$, along the $a$ and $c$ axis, up to pressures of 1.8 GPa. The pressure effect is strongly current direction dependent, indicating a significant anisotropy of the Fermi surface. For $I \parallel c$ $\rho(T)$ increases with pressure and develops a relative minimum at low temperatures ($T_{min} \sim 4.8$ K at 1.8 GPa) [5]. Here we concentrate on the low-temperature data obtained for $I \parallel a$.

A single-crystalline batch of $U_2Pt_2In$ was prepared by a modified mineralisation technique. Single-crystalline $U_2Pt_2In$ forms in the tetragonal $Zr_3Al_2$-type of structure (space group $P4_2/mnm$). The residual resistivity $\rho_0$ amounts to $\sim 115$ $\mu\Omega$cm for $I \parallel a$ [3]. With $\rho_{RT} \equiv \rho(300K) = 220$ $\mu\Omega$cm, a low residual resistance ratio $\rho_{RT}/\rho_0 = 1.9$ results. The resistivity was measured in the temperature interval 0.3-300 K, using a standard low-frequency four-probe ac-technique with a typical excitation current of ~ 100 μA. Pressures were exerted using a copper-beryllium clamp cell [4].

At ambient pressure the resistivity ($I \parallel a$) shows, upon cooling below 300 K, a weak maximum near 80 K, followed by a steady decrease at lower temperatures. Pressure leaves $r$(300K) unchanged, but results in an overall reduction of $r(T)$ below 300 K. In Fig.1 we show the low-temperature ($T<$ 3.2 K) data in a plot of $\Delta r \equiv r(T)-r_0$ versus $T^2$. The pressure data ($p \geq$ 0.2 GPa) show a steady reduction of $\Delta r$ with increasing pressure. Notice that the zero-pressure data have been measured in a different experimental set-up and are not in-line with this trend, as they fall in between the 0.2 and 0.6 GPa curves. This is possibly due to changes in the voltage contacts on the sample. The relative accuracy in the resistivity data for the different experiments amounts to ~10%, because of the poor determination of the geometrical factor.

A most important conclusion that can be drawn from the data presented in Fig.1, is the recovery of the Fermi-liquid (FL) $\Delta r \sim T^2$ law at moderate pressures. We have extracted $T_{FL}$, i.e. the upper temperature limit for the $\Delta r \sim T^2$ behaviour, by a least-squares fitting procedure. The pressure variation of $T_{FL}$ is shown in Fig.2. Within the error bars, the data are consistent with $T_{FL}$ being a linear function of pressure. Such a linear pressure dependence has been proposed by Rosch [6] for itinerant antiferromagnets in the paramagnetic regime close to a magnetic quantum critical point (QCP). Within this magnetotransport theory $T_{FL}$ is calculated as a function of the distance (measured by the pressure) to the QCP. $T_{FL}$ varies initially as $T_{FL} = a_1 (p-p_c)$ with a cross-over to $T_{FL} = a_2 (p-p_c)^{1/2}$ at higher distances, where $p_c$ is the pressure at the QCP. The pressure intervals in which the different laws are observed depend on the amount of disorder $x$ in the system ($x \approx$ 1/RRR = $r_0/r$(300K)). For $I \parallel a$, $x$ ~ 0.6, which indicates that our sample is in a regime of intermediate disorder. In this regime theory [6] predicts $T_{FL} = a_1 (p-p_c)$, while at the QCP $\Delta r \sim T^\alpha$ with the NFL exponent $a$ =1.5. The measured NFL exponent $a$ = 1.25±0.05 at $p_c$=0 [4], is rather close to the predicted value. Correspondingly, the coefficient $A$ of the FL $T^2$ term diverges upon approaching $p_c$. A strong increase of $A$ can be deduced from the data in Fig.1. A more detailed account of our high-pressure study can be found elsewhere [7].

The large residual resistivity value of our samples brings about the question whether the NFL behaviour in $U_2Pt_2In$ is due to Kondo disorder. However, the rapid recovery of the FL behaviour under pressure as probed by the resistivity data for $I \parallel a$ does not support this mechanism. Since the compressibility is isotropic [4], pressure is expected to result in the further broadening of the distribution of Kondo-temperatures and thus the concurrent NFL behavior would prevail.

In summary, we have measured the resistivity of the NFL material $U_2Pt_2In$ under pressure. For $I \parallel a$, the FL state is rapidly recovered. $T_{FL}$ is a linear function of the pressure, $T_{FL} \sim (p-p_c)$, with $p_c \approx 0$. This provides strong evidence for the location of $U_2Pt_2In$ at an antiferromagnetic quantum critical point.

**Acknowledgements:**

The authors acknowledge support obtained within the EC-TMR and ESF-FERLIN programmes. A. Matsushita is acknowledged for assistance in developing the pressure cell.

**Figure captions**

Fig.1  Resistivity under pressure measured for a current along the a-axis of single-crystalline $U_2Pt_2In$ in a plot of $\Delta r$ versus $T^2$.

Fig.2  $T_{FL}$ versus pressure for single-crystalline $U_2Pt_2In$ determined from resistivity data for $I \parallel a$. The solid line divides the non-Fermi liquid (NFL) regime from the FL regime. The linear behaviour is consistent with the presence of an antiferromagnetic quantum critical point at zero pressure.

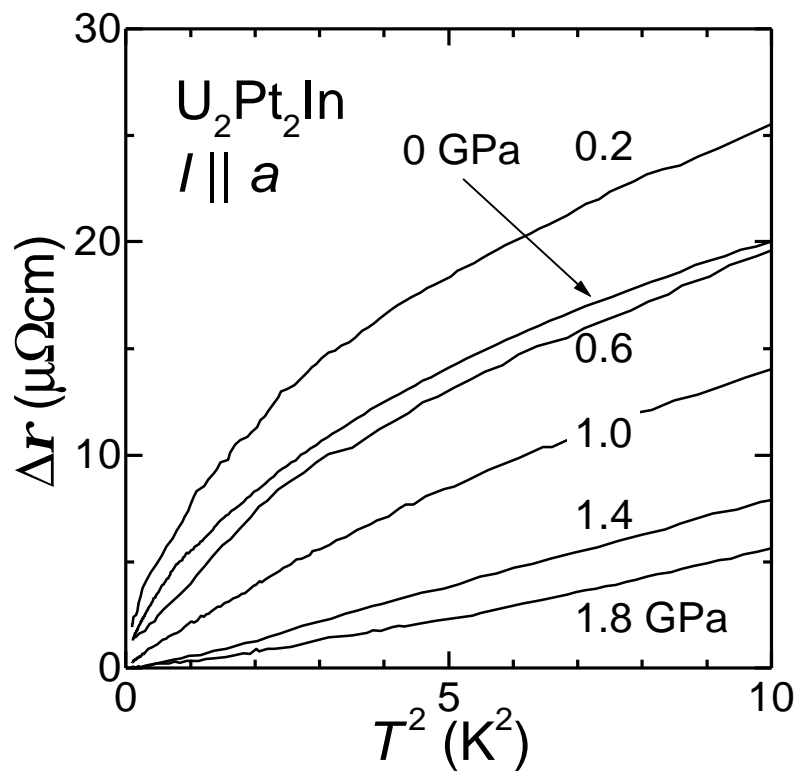

**Figure 1**

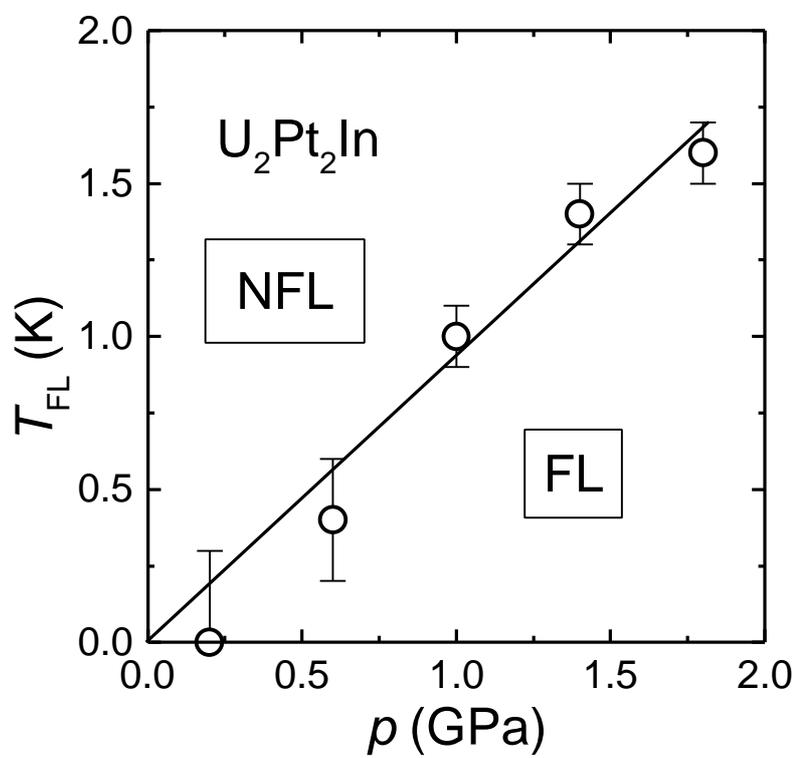

**Figure 2**